# Thermoelectric transport perpendicular to thin film heterostructures using Monte Carlo technique

Mona Zebarjadi and Ali Shakouri
Department of Electrical Engineering, University of California, Santa Cruz CA 95064
Keivan Esfarjani
Physics Department, Sharif U. of Technology, Tehran 11365-9161 Iran and
Physics Department, University of California, Santa Cruz, CA 95064

The Monte Carlo technique is used to calculate electrical as well as thermoelectric transport properties across thin film heterostructures. We study a thin InGaAsP barrier layer sandwiched between two InGaAs contact layers, when the barrier thickness is in 50nm-2000nm range. We found that with decreasing size, the effective Seebeck coefficient is increased substantially. The transition between pure ballistic thermionic transport and fully diffusive thermoelectric transport is also described.

## Introduction

The Seebeck coefficient[1] is usually referred to as a bulk material property. However, as the size of a material is decreased and the hot and cold junctions are closer to each other, the effect of interfaces becomes more important and the Seebeck coefficient is not independent of size anymore. For very thick films, transport is diffusive, and electrons have enough time to thermalize with the lattice. For thin films though, there is a ballistic contribution to transport as well. Some electrons pass through the thin film without experiencing any scattering. These don't have enough time to thermalize with the lattice. Consequently their energy remains high. The difference in the electron distribution in these two limits would result in different transport regimes: thermoelectric and thermionic.

This phenomenon is important in heterostructures where electrons experience different layers as they pass through the device. Such devices, based on the thermionic emission effect, can be used for refrigeration. They have been studied by different groups[2,3] and have attracted much interest in recent years both theoretically and experimentally[4,5,6]. Theoretical methods for analyzing electron transport in the structure depend on the length scale. For nanoscale sizes where quantum effects are important and transport is ballistic, quantum mechanical calculations such as Landauer formalism need to be implemented[7]. If the electron wavelength becomes smaller than the characteristic length of the structure, then the Boltzmann transport equation (BTE) can be an appropriate governing equation. Solving BTE, however is generally difficult even in its simplest form. For very long structures we are in the linear regime so the system obeys Ohm's law. But for micron and sub-micron scales, where most of heterostructure thermionic coolers work, transport is far from equilibrium and one needs to solve BTE beyond the linear response regime. Zheng and Chen attempted to solve BTE using the relaxation time approximation[8]. They found discontinuity of intensive parameters such as temperature and chemical potential at the junctions, on a scale compared to the barrier width. This is only correct if the barrier width is much larger than the inelastic length over which electrons and phonons can thermalize, and a local temperature and chemical potential can be defined. They furthermore predict an increase of the Seebeck coefficient as the width of the barrier is decreased. However, as width decreases (to the order of a mean free path), their results cease to be valid. This range is beyond the validity of their theory because the transport becomes increasingly ballistic.

Vashaee and Shakouri [9] solved energy and momentum balance equations. They also used the relaxation time approximation and included the electron-phonon interaction in a phenomenological way. Continuity of electron and phonon temperature along the structure was assumed and equilibration of electrons and phonons was investigated. These approximations can give an insight about the system behavior in different limits, but they are not rigorous since specific boundary condition for temperature and chemical potential are assumed.

A more exact solution of BTE can be obtained using the Monte Carlo (MC) simulation[10]. Several groups have used this powerful tool to study electric transport in heterostructures (see for example Damocles group works[11,12]). There were also a few attempts to study phonon transport using Monte Carlo simulation[13,14]; but there has been only one attempt to use MC to study thermoelectric effect in heterostructures[2]. Although this work can be considered as a good starting point, it is far from complete. First of all it is a bulk simulation, not a multilayer one. Secondly, carrier-carrier interaction is ignored, and finally phonon current and lattice temperature have not been extracted from the simulation.

In this paper, we use MC simulation to study electron transport and calculate the phonon current. The objective of this work is to study thermoelectric and thermionic transport perpendicular to thin film heterostructures. Furthermore we use this knowledge to investigate size effects on transport. One should note that there is a lot of work on low dimensional thermoelectrics (see for example Hicks and Dresselhaus[15]). However, there are no calculations on non-equilibrium effects due to close proximity of cold and hot junctions.

This paper is organized as follows: First the model and theory are explained. Then we introduce the particular single barrier system we are investigating and discuss heat and charge propagation along it. The contributions of Peltier cooling and non-uniform joule heating as well as phonon thermal current will be discussed. The size of the regions where thermoelectric cooling and heating happens is calculated. Subsequently, we investigate how thermoelectric properties of the material changes as we go from thermionic limit where transport is ballistic to thermoelectric limit where transport is mainly diffusive. Finally some general conclusions are given in the last section.

## Model and Theory

Consider a layer of InGaAsP sandwiched between two layers of InGaAs, used as contact layers to the cathode and anode as shown in FIG 1. The contact layers are long enough that electrons reach local equilibrium before entering the main layer. This is necessary because otherwise boundary effects in contact-electrode junctions would affect the results. To make the potential drop happen over the main layer, doping of contact layers is taken to be high compared to the main layer.

We implemented the MC technique to simulate electrical transport in a single barrier heterostructure. The model is solved in three dimensions both in k and r space. A non-parabolic multi-valley band structure is used for each material. Scattering mechanisms considered in the simulation include: intra-valley polar optical and acoustic phonons, inter-valley optical phonons, and ionized impurity scattering[16]. Acoustic phonon scatterings are considered as inelastic processes. Because electron scattering rate from acoustic phonons is much smaller than that of optical phonons in room temperature, exchange energy between electrons and acoustic phonons can be ignored. Electron-electron interaction is treated using one-dimensional Poisson solver coupled with MC simulation. By solving the Poisson equation after each time step, the electric field along the system can be obtained. This electric field contains both the external field and the electron-electron interaction in itself, so it can be used as the force acting on carriers for the next time step in the simulation.

In lateral directions (x and y) periodic boundary condition is used at the two ends of the sample. In normal direction (z) injection takes place. Electrons are injected through the contact-electrodes junction using the Fermi distribution of the same material as the contact layer. This kind of injection lets the electrons reach equilibrium as fast as possible, but there are still non-equilibrium regions for the injected electrons. In the final analysis, these regions are disregarded in order to remove the boundary effects (FIG 2).

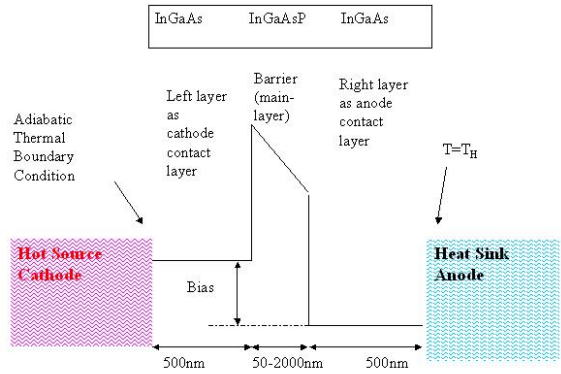

**FIG 1. Schematic of the heterostructure**

Tunneling current is not included in the simulation, as the main layer is thick enough to block this current. Also, applied voltage is not very high and field emission can be neglected. Conservation of lateral momentum and total energy is implemented for electrons, which pass the heterojunction.[17]

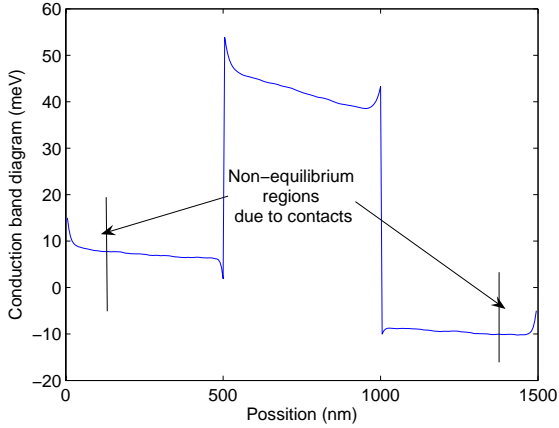

**FIG 2. Conduction band Profile along the heterostructure.**

In order to understand heat transport one needs to consider both phonons and electrons. In MC simulation, phonons are only present through electron-phonon scattering. Simulation of phonon current with electron current is possible but it is very time consuming. An alternative way is the following: phonons exchange energy with electrons through electron-phonon scattering. This provides an energy source for phonons. During MC simulation we can calculate the energy transferred from electrons to phonons versus time and position. This energy can be used as a source term ($\delta(z)$) to create phonon current ($Q_{ph}$).

$$\nabla \cdot Q_{ph} = \delta(z) \quad (1)$$

Using the thermal conductivity of phonons ($\kappa_{ph}$) - from experimental results-, lattice temperature (T) along the sample can be calculated through integration of:

$$Q_{ph} = -\kappa_{ph} \nabla T \quad (2)$$

With the following boundary conditions: a heat sink at the anode end (T=300K) and zero thermal current at the cathode end ($Q_{ph}$=0). We are mainly interested in low field transport which is important in thermoelectric and solid state thermionic applications. In this case, hot phonon effect can be ignored and Fourier heat diffusion law is a good approximation.

The program was tested for bulk GaAs. We obtained a mobility of 8350 cm$^2$ V$^{-1}$s$^{-1}$ at room temperature and a saturation velocity of $1.2 \times 10^7$ Cm/s which are in good agreement with experimental data[18].

Table I- Material parameters for simulation, (all parameters are taken from Ioffe[18] or Landolt-Börnstein online database[19].

| Parameter | Unit | $In_{1-x}Ga_xAs$ | $In_{1-x}Ga_xAs_yP_{1-y}$ |
|---|---|---|---|
| x and y (mole fractions) |  | x= 0.47 | y= 0.8<br>x= 0.1896y/(0.4176-.0125y) |
| $E_c$ conduction band measured from GaAs conduction band | eV | -.7(1.4240-.324-.7x -.4$x^2$) | -.268y-.003$y^2$ |
| Effective mass (Γ valley) |  | (0.023+0.037x+.003$x^2$)$m_0$ | (0.077-0.050y+0.014$y^2$)$m_0$ |
| dielectric constant |  | 15.1-0.87x+0.67$x^2$ | 12.5+1.44y |
| Density | gr/$cm^3$ | 5.68-0.37x | 4.81+0.552y+0.138$y^2$ |
| Sound velocity | $10^5$ Cm/s | 4.28+0.96x | 5.092+0.355y |
| Thermal conductivity | W/mK | 6 | 4 |
| Polar optical phonon energy | eV | 0.03536x+0.030(1-x) | 0.0424 |
| Acoustic deformation potential (Γ valley) | eV | 5.89 | 8.3-5.4y+2.8 $y^2$ ** |
| Non-parabolicity | $eV^{-1}$ | 1.307 | 0.968 ** |
| Energy separation Γ- L valley | eV | 0.553 | 0.323 |
| Energy separation Γ- X valley | eV | 0.668 | 0.457 |

** measured for y=0.5

## Results and discussions

### Heat propagation along the heterostructure

The simulated heterostructure consists of InGaAs/InGaAsP/InGaAs. Some of the important material parameters are listed in Table I. All the results are at room temperature. The heterostructure area is 0.5×0.5μ$m^2$. Contact layers are 0.5 μm long and the main layer varies from 50nm to 1 μm.

Here we give a discussion of heat transport along the structure from the simulation results. FIG 3 shows the electron to phonon heat exchange (called *source term*) profile in contact layers and main layer as a function of position for a 500nm long layer. Peltier cooling occurs just before the barrier (negative source term, left layer) while hot electrons leave the cathode contact layer. The remaining cold electrons try to rebuild the Fermi distribution. These cold electrons gain their needed energy from phonons, so the junction is cooled down(s<0 at location z=0.75um). When electrons enter the barrier, they lose their kinetic energy by the band offset value. Because their energy is now below $k_B T$, they need to gain energy mostly from the electric field but also in part from phonons. This happens over a small length (of the order of nm) at the beginning of the barrier. As soon as electrons gain enough energy, they start transferring their additional gained energy to phonons (joule heating). The small heating peak at the beginning of the main layer is due to band bending and the large slope of the potential at that location. Electrons reach equilibrium with phonons within an energy relaxation length ($l_E \approx 150$ nm), after which the energy that electrons gain from the external field is equal to the energy that they transfer to phonons and the electron-phonon energy exchange profile is almost flat.

At the junction between the barrier and the right layer (at z=1.25um), Peltier heating occurs. This time as electrons enter the anode side their kinetic energy suddenly increases by the band offset value. They dissipate their extra energy inside the anode layer. Both the peak of Peltier heating and the dip of Peltier cooling are shown in figure 3. In this way, electrons transfer heat from the cathode side to the anode side.

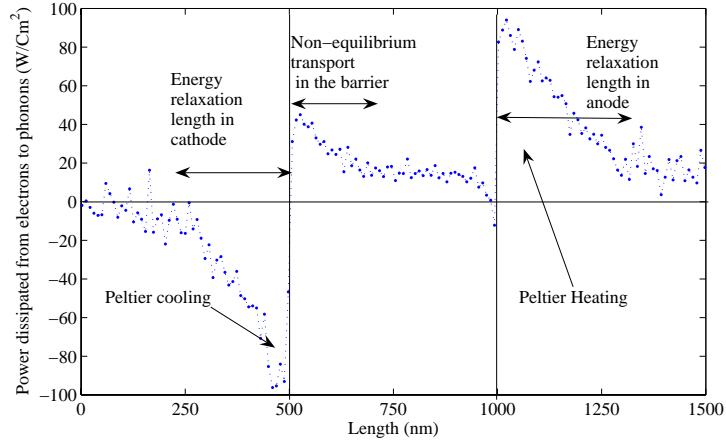

**FIG 3. Source term (electron-phonon energy exchange) along the heterostructure obtained from MC simulation. The applied bias is 40 mV. Non-equilibrium regions at the ends of the contacts are disregarded as explained before. Peltier cooling at the cathode-barrier junction and Peltier heating at the anode-barrier junction can be observed.**

Energy relaxation lengths in different regions (cathode, barrier and anode) depend strongly on electron-phonon coupling. Polar optical phonons are mainly important because most of the energy exchange is due to this type of phonons. In the case of strong electron-phonon coupling, the energy relaxation length, over which Peltier cooling (or heating) occurs, becomes small. In steady state, as the rate of energy transferred from phonons to electrons is equal to the rate of energy transported by electrons, if the relaxation length is smaller, then the depth of $\delta(z)$ should be larger so that its integral remains the same. Therefore, the depth of $\delta(z)$ is an increasing function of the electron-phonon coupling, whereas the energy relaxation length is a decreasing function of it. The energy relaxation length is typically an order of magnitude larger than the momentum relaxation length, which gives the electron mobility in the material.

FIG 4 shows the calculated temperature profile for the same structure. In this structure thermal conductivity for both materials is of the same order. To prevent a temperature drop over the right layer, one should use a material with a high thermal conductivity in comparison with the barrier layer. In this simulation though, we are just interested in the thermal properties of the barrier. The temperature drop over the barrier just depends on the thermal conductivity of that layer.

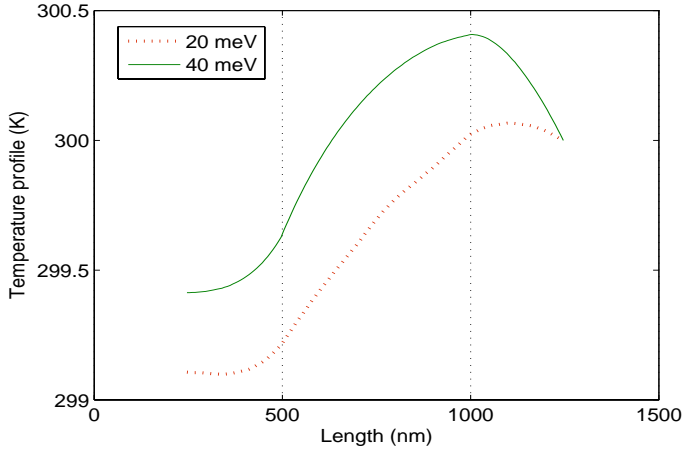

**FIG 4. - Temperature profile along the heterostructure for two different voltages**

### Size dependence of maximum cooling
Next, we explain how thermoelectric properties change as we decrease the barrier size to sub-micron scale.

The parameter that determines the thermoelectric energy conversion efficiency of different materials is the figure-of-merit $ZT = \dfrac{\sigma S^2 T}{\kappa}$.

S is the effective Seebeck coefficient, σ is the effective electrical conductivity, and κ is the thermal conductivity.

Conductivity is expected to be constant if there is no band bending. But in a real physical device band bending occurs at the junctions (see FIG 2) and it adds to its resistance. The effect of band bending becomes more important as we go to shorter barriers. FIG 5 shows the dependence of the main layer resistivity as a function of length.

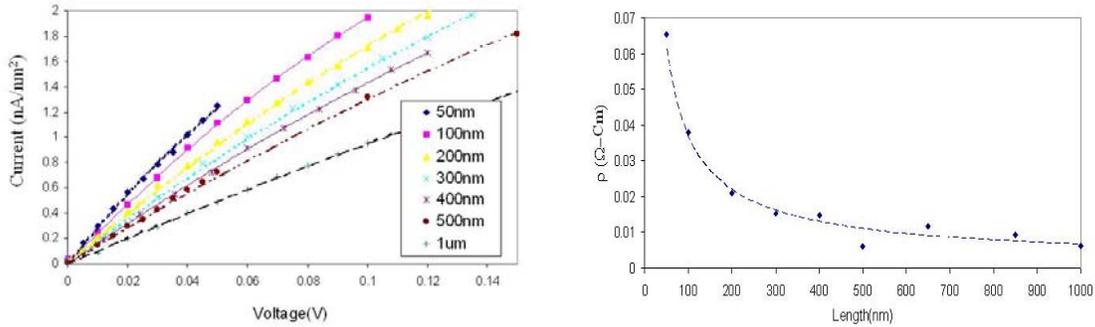

**FIG 5- a) I-V curves for different barrier lengths, obtained form MC simulation. Each of these curves is fitted with a second order polynomial to obtain resistivity. b) Resistivity versus layer thickness for different structures.**

The next important parameter is the effective Seebeck coefficient which increases the efficiency. In order to find the Seebeck coefficient from the MC simulation, we can use the energy balance equation inside the barrier.

$$Q_{TI} = STj - \alpha Vj - \frac{\kappa}{d}\Delta T \quad (3)$$

$$Q_{TI} = \text{heat load on cathode side} = Q_{ph} + Q_e \mid_{z=z0}$$

The first term (*STj*) is the Peltier cooling. S is the effective Seebeck coefficient, T is temperature and *j* is the current density per unit area. The second term is the amount of Joule heating that returns to the cathode. Here V is the voltage drop over the barrier; *Vj* is the joule heating inside the barrier and α is the percent of it returning to the cathode side. Finally the last term is the phonon thermal current. In this term ΔT is the temperature drop over the barrier and d is the length of the barrier [2]. $Q_{TI}$ is the total thermal current coming inside the barrier at the barrier-cathode junction (z=0.5 um). $Q_{TI}$ and ΔT as a function of current can be obtained directly from the MC simulation. Fitting these quantities versus current, S can be extracted from the linear term because *STj* is the only linear term in equation 3. $Q_{TI,}$ ΔT and S are shown in FIG 6, 7 and 8 respectively.

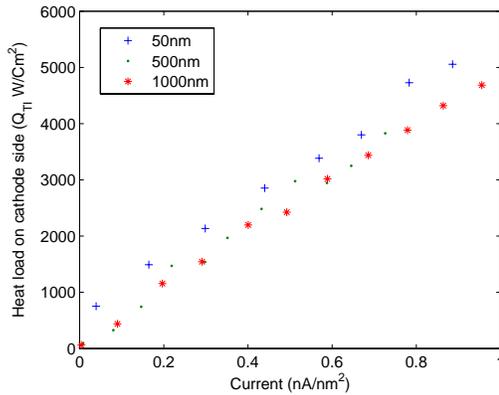

**FIG 6- Heat load at the cathode-barrier junction as a function of current for different barrier sizes.**

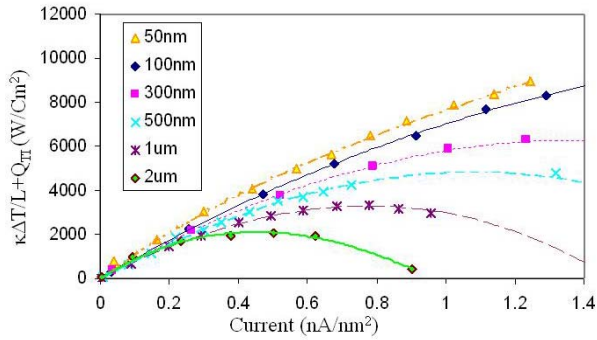

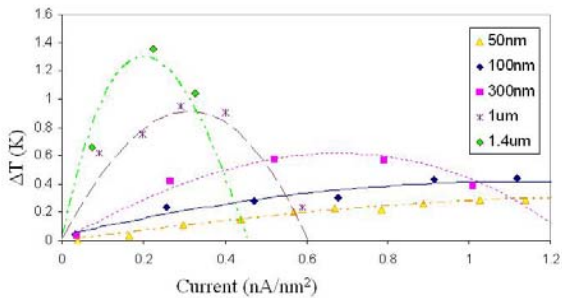

**FIG 7-a) Cooling of electron gas at the cathode/barrier junction as a function of current. The cooling at this junction creates a temperature gradient across the barrier and it removes joule heating in the**

**cathode. b) Temperature difference across barrier versus current for different barrier thicknesses. Symbols are MC simulation data and lines are fits for each set of data. Seebeck coefficient can be obtained from the linear fit of figure a.**

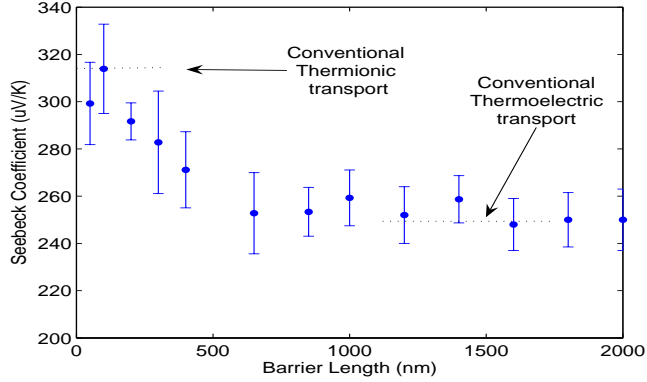

**FIG 8- Seebeck coefficient for InGaAsP versus layer thickness, obtained by fitting MC results with the energy balance equation (equation 3)**

It can be seen from FIG 8 that with decreasing length, the Seebeck coefficient is increased by 60μV/K approximately. This increase was expected as we discussed before. The Seebeck coefficient can be referred to as the average heat transported per unit charge divided by temperature. For very short barriers transport is completely ballistic and the energy of electrons passing the barrier is determined by the barrier height and electron distribution in the cathode and anode sides. We see a ballistic limit of ~315 μV/K in Fig. 8 that matches approximately the simple thermionic estimate.

$$S = \frac{K_B}{e}(\frac{E_{barrier}}{K_B T} + 2) \sim 344 \mu V/K, \text{ where } E_{barrier} = 50 \text{ meV}$$

If we increase the barrier thickness slightly (FIG 3) electrons have enough time to undergo several scatterings and gains energy. So in this range, electrons pass the barrier with higher energies. Because the quasi-Fermi level is below the conduction band inside the barrier, the amount of heat per charge is higher, as is the Seebeck coefficient. For longer barriers (>100nm), the extra energy is transferred to phonons and electrons and phonons are thermalized. The average energy earned by electrons is lower, that leads to a lower thermopower. Thermopower converges to the bulk value in the limit of thick barriers >500 nm. We see a diffusive limit of ~ 255μV/K in Fig.8. This is approximately equal to the bulk thermopower which is 295 μV/K. The difference is due to ignoring Pauli exclusion principle in our calculations.

$$S = \frac{1}{eT}(\frac{\int v^2(E-\mu)\tau_E Dos(E)\frac{\partial f}{\partial E}dE}{\int v^2 \tau_E Dos(E)\frac{\partial f}{\partial E}dE}) \sim 295 \mu V/K$$

Inclusion of Pauli exclusion principle will be the subject of a future publication.

## Summary and conclusion

In summary, we developed a Monte Carlo code to study thermoelectric transport perpendicular to thin film heterostructure sandwiched between two contact layers. The size of the regions where thermoelectric cooling and heating happens near the junctions is identified. This size is related to energy relaxation length, which decreases with the electron-phonon coupling strength. One can therefore conclude that in samples with weak electron-phonon coupling where the size of non-equilibrium regions is increased, one is closer to the thermionic limit and the Seebeck coefficient is larger. It is interesting that the thermoelectric cooling and heating regions are contained in the highly doped contact layers. Thus additional thermal boundary contact resistance between cathode and anode layers and the thermoelectric leg could improve the device performance if they do not affect electron transport. Furthermore by considering non-uniform Joule heating, which occurs due to non-equilibrium transport of electrons from the cathode to the barrier, thermoelectric properties of short legs is calculated. With decreasing length, we have observed a higher Seebeck coefficient. For the considered structure, an increase of 60 μV/K was found. This increase in the Seebeck coefficient is due to high-energy-electron filtering of the barrier. An energy relaxation length larger than the barrier width, allows electrons to keep their high energy and not relax during the barrier traversal.

## Acknowledgements

We are grateful to Dr. Ceyhun Bulutay for his valuable insights and especially his help on developing the code for bulk material. M.Z. would like to thank Dr. Bilal Tanatar for his support and the hospitality of Bilkent University where part of this work was done. We would also like to thank Dr. Daryoosh Vashaee for our useful discussions.

[1] H.J. Goldsmid, *Thermoelectric refrigeration* (Plenum Press, New York, 1964)

[2] A. Shakouri, E. Y. Lee, D. L. Smith, V. Narayanamurti, and J.E. Bowers, Micro-scale Thermophys. Engineering **2**, 37 (1998).

[3] G.D. Mahan and L.M. Woods, Phys. Rev. Lett. **80**, 4016 (1998).

[4] A. Shakouri, C. LaBounty, J. Piprek, P. Abraham and JE. Bowers, Appl. Phys. Lett. **74**, 88 (1999).

[5] X. Fan et al, Appl. Phys. Lett. **78**, 1580 (2001).

[6] JH Zhang, NG Anderson, and KM Lau, J. Appl. Phys. **83**, 374 (2003).

[7] K. Esfarjani, M. Zebarjadi, and Y. Kawazoe, Phys. Rev. B **73**, 085406 (2006).

[8] TF. Zeng and G. Chen, J. Appl. Phys. **92**, 3152 (2002).

[9] D.Vashaee and A. Shakouri, Microscale Thermophysical Eng. 8, **91** (2004).

[10] C. Jacoboni and L. Reggiani, Rev. Mod. Phys. **55**, 654 (1983).

[11] M. V. Fischetti and S. E. Laux, Phys. Rev. B **38**, 9721 (1988)

[12] S. E. Laux, M. V. Fischetti and D. J. Frank, IBM J. Res. Develop. **34**, 4 (1990).

[13] R.B. Peterson, J. Heat Transfer **116**, 815 (1994)

[14] S. Mazumder and A. Majumdar, J. Heat Transfer **123**, 749 (2001).

[15] L.D. Hicks and M.S. Dresselhaus, Phys. Rev. B **47**, 12727 (1993).

[16] M. Lundstrom, *Fundamental of carrier transport* (Cambridge Universiy press, 2000)

[17] P. Garcias-Salva, et al, Microelectronic Engineering **51-52**, 415 (2000).

[18] Ioffe Physico-Technical Institute; http://www.ioffe.rssi.ru/SVA/

[19] Landolt-Börnstein online data base;
http://www.springer.com/west/home/laboe?SGWID=4-10113-0-0-0

[20] Landolt-Börnstein online data base;
http://www.springer.com/west/home/laboe?SGWID=4-10113-0-0-0